# Robust Control for Uncertain Air-to-Air Missile Systems

**Joshua Farrington, Xinhua Wang**
Aerospace Engineering,
University of Nottingham, UK
Email: wangxinhua04@gmai.com

**ABSTRACT**
Air-to-air missiles are used on many modern military combat aircraft for self-defence. It is imperative for the pilots using the weapons that the missiles hit their target first time. The important goals for a missile control system to achieve are minimising the time constant, overshoot, and settling time of the missile dynamics. The combination of high angles of attack, time-varying mass, thrust, and centre of gravity, actuator delay, and signal noise create a highly non-linear dynamic system with many uncertainties that is extremely challenging to control. A robust control system based on saturated sliding mode control is proposed to overcome the time-varying parameters and non-linearities. A lag compensator is designed to overcome actuator delay. A second-order filter is selected to reduce high-frequency measurement noise. When combined, the proposed solutions can make the system stable despite the existence of changing mass, centre of gravity, thrust, and sensor noise. The system was evaluated for desired pitch angles of 0° to 90°. The time constant for the system stayed below 0.27s for all conditions, with satisfactory performance for both settling time and overshoot.

## NOMENCLATURE

| Symbol | Name | Unit |
|---|---|---|
| $\alpha$ | Angle of attack | $rad$ |
| $\bar{q}$ | Dynamic pressure | $Pa$ |
| $C_{m\alpha_0}$ | Pitch moment coefficient due to angle of attack | - |
| $C_{m\delta T_0}$ | Pitch moment coefficient due to tail fin deflection | - |
| $C_{z\alpha_0}$ | Pitch force coefficient due to angle of attack | - |
| $C_{z\delta T_0}$ | Pitch force coefficient due to tail fin deflection | - |
| $I_{yy}$ | Pitch moment of inertia | $kg\,m^2$ |
| $\rho$ | Air density | $kg\,m^{-3}$ |
| $Ma$ | Mach number | - |
| $\theta$ | Pitch angle | $rad$ |
| $x_{CG}$ | Centre of gravity from nose | $m$ |
| $d$ | Missile diameter | $m$ |
| $l$ | Missile length | $m$ |
| $m_0$ | Initial mass | $kg$ |
| $q$ | Pitch rate | $rad\,s^{-1}$ |
| $S_{ref}$ | Reference area | $m^2$ |
| $T_0$ | Initial thrust | $N$ |
| $V_M$ | Missile velocity | $m\,s^{-1}$ |

## 1. INTRODUCTION

Agile air-to-air missiles are currently in use on modern military aircraft platforms such as the Eurofighter Typhoon and the Lockheed Martin F-35 II. Air-to-air missiles operate over a large flight envelope and, as such, there are many uncertainties in their flight profile.
Missiles are inherently unstable to aid with manoeuvrability, however, there is the need to control them so that they hit their intended target. The jobs of the control system are to make the missile dynamics stable and to implement the intended flight mission.

This paper expands upon existing robust controllers to overcome the unknown, time-varying uncertainties with the benefit of faster response times for agile missiles. The goals for such a missile are taken from Reichert [1] and are a time constant of less than 0.35s and a steady-state error of less than 5%, where the time constant ($\tau$) is the time taken for the signal to reach $1-\frac{1}{e} \cong 63.2\%$ of the required step-input.

## 2. BACKGROUND & LITERATURE

### 2.1 MISSILE AUTOPILOT

Missile guidance navigation and control autopilot have the typical configuration described in Figure 1, whereby the missile's sensors give target information to the guidance loop, which in turn gives an acceleration command to the control loop.

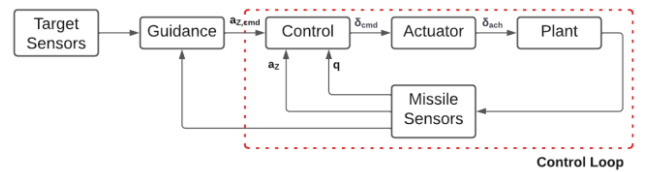

**Figure 1: Typical missile autopilot configuration [2]**

**Table 1: Summary of Missile System Problems and Solutions.**

| Technical Problems | Current Methods | Method Description |
|---|---|---|
| Unstable static and dynamic characteristics | Sliding mode controller with linear-quadratic regulator [3] | <ul><li>Does not require knowledge of system non-linearities and uncertainties for controller design.</li><li>The simplified system properties of unvarying mass in Bużantowicz's paper neglect challenging control regimes.</li><li>Drawback of chattering existing in the control output.</li></ul> |
| | Non-linear adaptive sliding mode control [4] | <ul><li>Non-linear system model possible.</li><li>Actuators are not considered as part of the system, unlike in Bużantowicz's paper. Consideration of the actuators is essential as they introduce performance penalties for sliding mode control systems.</li><li>Drawback of the method is that it is only fit for control systems with parametric uncertainties.</li></ul> |
| Changing mass, thrust & centre of gravity | H∞ control [5] | <ul><li>Functions over large Mach variation.</li><li>Works for non-linear systems.</li><li>Has better performance than back-stepping control [5].</li><li>The paper by Mahmood does not consider actuator constraints, leading to satisfactory controller performance but with nozzle deflections of approximately 65° which is too high for most missiles [6].</li><li>Drawbacks of the method are that it is only fit for control systems with small system errors, and it is only optimal against a cost function instead of regular controller parameters such as settling time.</li></ul> |
| | Robust gain-scheduling [7, 8] | <ul><li>Low computational cost [7].</li><li>Functions over varying thrust, centre of gravity, and mass.</li><li>Only the boost phase of flight is considered compared to the full flight in Sankar [2].</li><li>One benefit is that the actuators perform within regular constraints of angle when compared to Mahmood [5].</li><li>Drawbacks are that the switching between different gains cannot cover all system scenarios, and it is only suitable for parametric systems.</li></ul> |
| | Linear parameter varying [9] | <ul><li>Works for non-linear systems, however it requires a linear approximation of the system.</li><li>Performs well with actuators modelled and within specified actuator requirements.</li><li>±10% simulated for each flight parameter so works within a boost phase of flight.</li><li>Flight envelope considered is only the short boost phase [9] rather than the whole flight envelope of Sankar [2].</li></ul> |
| High angle of attack | High order sliding mode control [2] | <ul><li>Non-linear system modelled.</li><li>Sliding mode control provides output to the guidance loop.</li><li>Second-order non-linear actuators are considered in the design.</li><li>Boost phase of flight not considered where parameters vary rapidly.</li><li>Drawback is a high computational cost and chattering still exists in the controller.</li></ul> |

The relationship between commanded normal acceleration and the pitch rate ($\dot{\theta}$) is known to be [10]:

$$a_{z,cmd} = \frac{\dot{\theta}}{V_M} \quad (1)$$

Therefore, for an assumed constant missile velocity, when modelling the longitudinal dynamics, the controller can be designed based upon pitch angle ($\theta$) or pitch rate ($\dot{\theta}$). This paper uses the control loop part of the autopilot structure to design a pitch control system.

## 2.2 TECHNICAL CONTROL PROBLEMS

The uncertainties in the missile system are due to the changes in aerodynamic characteristics. These cause the system to be statically and dynamically unstable, as well as the entire system being highly non-linear. This presents significant control challenges when the system is not fully modelled with the unknowns. The objective is to design a control system in such a way that it can overcome the combination of the uncertainties, as well as any additional disturbances to the system, such as signal noise and wind, without knowledge of the exact values.

### 2.2.1 EFFECT OF DYNAMIC PRESSURE

The missile can be launched from a range of altitudes and speeds and, as the properties of air change dependent on altitude for example air density, this affects the amount of dynamic pressure that the missile experiences (see Eq.(2)).

$$\bar{q} = \frac{1}{2}\rho V_M^2 \quad (2)$$

The missile coefficients of lift and drag are dependent on the dynamic pressure, so as it changes, so do the forces the missile experiences.

### 2.2.2 EFFECT OF MASS & CENTRE OF GRAVITY

As the missile burns fuel, the mass rapidly reduces, and the centre of gravity moves forwards from its initial position [7]. If the mass changes such that the centre of gravity is behind the centre of lift then the dynamics are statically unstable, i.e., if left uncontrolled then any perturbation to the missile will be amplified instead of corrected by the natural dynamics. This is a tricky control scenario due to the lack of self-correcting behaviour, and the controller needs to work for both the stable and unstable control regions.

### 2.2.3 EFFECT OF ANGLE OF ATTACK

The functions of missile aerodynamic coefficients include the angle of attack (AOA, see Figure 3). Agile missiles conduct rapid manoeuvres as commanded by the autopilot, resulting in rapid changes in, and high angles of, the AOA. This is difficult for controllers as the lift-AOA curve for aerofoils becomes non-linear, which in turn makes the resultant dynamic system highly non-linear due to its dependence on AOA (Eq. (11) and Eq. (12)).

## 2.3 CURRENT MISSILE CONTROL METHODS

Presently the method of designing the control systems is ad hoc gain scheduling. This method divides up the flight envelope into different operating regions, an adequate linear controller is designed for each region, the control gains are then interpolated between the regions [9, 11].

There are two drawbacks to this approach: the models are linear whereas the equations of motion for a missile are highly non-linear, especially during high AOA manoeuvres where small angle approximations will not hold; there is the assumption that the scheduled variable should vary slowly [12] which is not the case for a missile. This is particularly noticeable during the initial boost phase of flight where mass properties change extremely rapidly [7, 8]. Ad hoc gain scheduling is therefore a far from ideal control method. Current research is looking into robust control techniques to overcome the problems outlined in Section 2.2 (see Table 1).

## 2.4 CONTROL STRATEGY FOR UNCERTAIN MISSILE FLIGHT DYNAMICS

The most popular controller is the PID controller, however it is an extremely limited method for controlling systems. The missile system has numerous non-linear and unknown terms, therefore the PID controller cannot be used to control this system as the disturbances would cause the system to become unstable. Robust controllers can completely reject disturbances and modelling errors, successfully tending to stability in finite time [13]. A robust controller that can overcome the errors under all circumstances is therefore highly desirable.

At present, there are several robust controller types proposed in research which have been summarised in Table 1 on the previous page.

The saturated sliding mode controller (SMC) has been chosen for this paper as it does not require knowledge of the system's nonlinearities and uncertainties for the design of the controller.

SMC systems are designed to drive and constrain the system state to the selected sliding surface, denoted as $s$. Typically, an SMC will consist of a switching function that forces the system back onto the sliding surface, and an equivalent control law that will make the switching function tend towards the stable equilibrium condition [14]. Figure 2 illustrates how, from initial conditions $[y_0 \quad \dot{y}_0]$, there is a reaching phase (RM) where the system is driven along the sliding manifold (s), then the switching function drives the system in a sliding mode (SM) towards the origin, along the line $s = 0$.

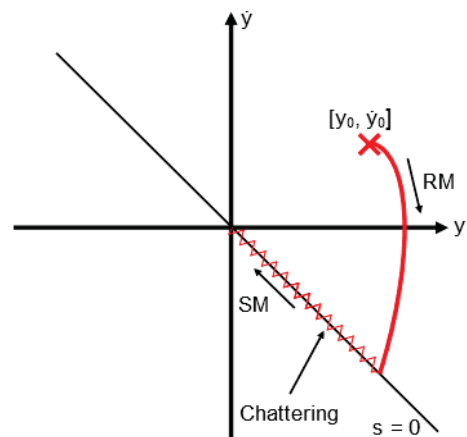

**Figure 2: Sliding mode control concept [15].**

The main advantages of a sliding mode controller are the dynamic behaviour of the system can be controlled through the choice of the switching function [14]; the controller can be used for non-linear systems; uncertainty is included as part of the design process, which gives additional robustness properties compared to usual feedback control techniques. The disadvantage is the chattering phenomenon (see Figure 2) caused by the switching function about the line $s = 0$, which can cause actuator damage. This is discussed further in Section 5.1.1.

The SMC technique selected for this paper was the SMC based on bounded uncertainties from Liu [14] as a non-linear plant can be defined, and the changing dynamics modelled. This can be described as the following non-linear system:

$$\ddot{\theta}(t) = f(t) + \Delta f(t) + g(t)u(t) + \Delta g(t)u(t) + d(t) \quad (3)$$

Where $\theta$ is the measured output, $f$ and $g$ are known non-linear functions, $\Delta$ represents the unknown dynamics, $u$ is the control torque, and $d$ is the unknown disturbance. The unknown terms can be grouped as:

$$D(t) = \Delta f(t) + \Delta g(t)u(t) + d(t) \quad (4)$$

Hence Eq. (3) can be written in terms of knowns and unknowns as:

$$\ddot{\theta}(t) = f(t) + g(t)u(t) + D(t) \quad (5)$$

In practice, we can obtain the upper bounds of these parameters. A controller can then be selected in the form:

$$u = \frac{1}{g}\left[-f + \ddot{\theta}_d + c\dot{e} + \eta \mathrm{sgn}(s)\right] \quad (6)$$

Where $\ddot{\theta}_d$ is the desired value, $c$ and $\eta$ are controller parameters to be chosen, and $s$ is the sliding surface. It can be shown that the resulting system is autonomously stable if:

$$\eta \geq |D(t)| \quad (7)$$

For more details on SMC see Liu [14]. This means to design the controller a model of the known system dynamics is needed (Section 3), and of the time-varying dynamics (Section 4).

## 3. MISSILE SYSTEM DYNAMIC DESCRIPTION

The missile to be controlled has the configuration shown in Figure 3, whereby the tail is deflected by angle $\delta_T$ and the nozzle deflected by angle $\delta_N$ to control the pitch of the missile.

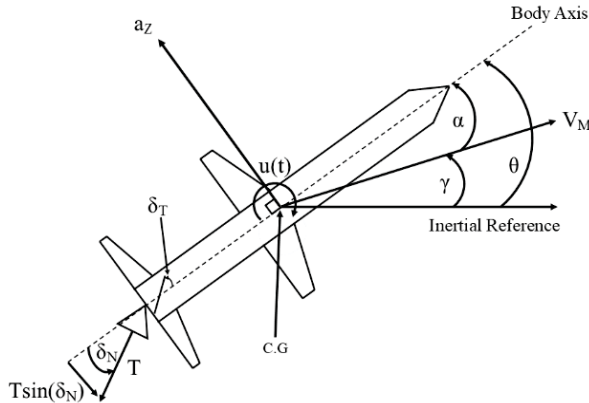

**Figure 3: Missile pitch dynamics, adapted from [16].**

The missile is based upon parameters from Mracek [17] and the characteristics are described in Table 2.

**Table 2: Missile Parameters**

| Variable | Value | Unit |
|---|---|---|
| $V_m$ | 1,021 | $m\,s^{-1}$ |
| $m_0$ | 162 | $kg$ |
| $I_{yy}$ | 187 | $kg\,m^2$ |
| $C_{m\alpha_0}$ | -5.5313 | - |
| $C_{m\delta T_0}$ | -6.6013 | - |
| $C_{z\alpha_0}$ | -1.2713 | - |
| $C_{z\delta T_0}$ | -7.5368 | - |
| $\bar{q}$ | 638.5 | $kPa$ |
| $S_{ref}$ | 0.0507 | $m^2$ |
| $d$ | 0.2540 | $m$ |
| $l$ | 3.72 | $m$ |

Missiles have sensors to detect the AOA ($\alpha$) and pitch rate ($q$), such as an AOA vane and inertial measurement unit respectively. Mracek shows that the missile's longitudinal equations of motion can be written in state-space notation [17], which is useful for modelling the system in MATLAB Simulink. The resulting equations of motion are therefore modelled in the following form:

$$\begin{aligned}\dot{x} &= Ax + Bu \\ \dot{y} &= Cx + Du\end{aligned} \quad (8)$$

Where:

$$x = [\alpha \quad q]^T \quad (9)$$

$$u = [\delta_N \quad \delta_T]^T \quad (10)$$

The real-time values for time-varying parameters will be used, making the system non-linear.

The equations of motion are derived from Figure 3. The controller will be based on the pitch, therefore the uncertain dynamics ($\Delta$) also need to be derived for this equation so that suitable controller parameters can be selected:

$$\dot{\alpha} = \left[\frac{QS_{ref}C_{z\alpha}}{mV_M}\alpha + q + \frac{T_0}{mV_M}\delta_N + \frac{QS_{ref}C_{z\delta_T}}{mV_M}\delta_T\right] \quad (11)$$

$$\dot{q} = \left[\frac{QS_{ref}dC_{m\alpha}}{I_{yy}}\alpha + \frac{T_0(l-x_{CG})}{I_{yy}}\delta_N + \frac{QS_{ref}dC_{m\delta_T}}{I_{yy}}\delta_T\right]$$
$$+ \left[\frac{QS_{ref}d\Delta C_{m\alpha}}{I_{yy}}\alpha - \frac{(T_0+\Delta T_0)\Delta x_{CG}}{I_{yy}}\delta_N\right.$$
$$\left.+ \frac{(T_0+\Delta T_0)}{I_{yy}}(l\Delta\delta_N - x_{CG}\Delta\delta_N - \Delta x_{CG}\Delta\delta_N) + f(\Delta m_0)\right]$$
$$(12)$$

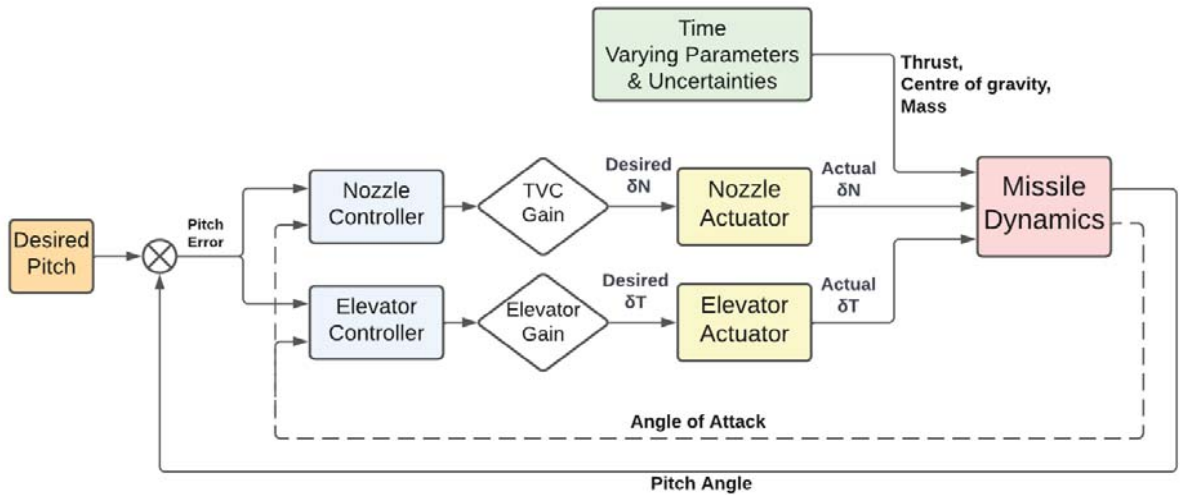
**Figure 4: Missile control loop.**

The equations can be verified from literature: Mracek and Ridgely derive equations for an elevator controlled missile [17], derivations in Lee et al. for a thrust vector controlled missile show that, in addition to these equations, there needs to be a nozzle term ($\delta_N$) [18].

The following assumptions were made when deriving the equations of motion:
- The only rotation is around the pitch (y-y) axis for the 1 degree-of-freedom model.
- The tail is the only lifting force.
- The missile travels at a constant velocity.
- The missile is a rigid body with no flexing therefore the sensor measurements do not need to be corrected for flex.
- The nozzle deflection is a small angle ($\leq 15°$) [6].

## 4. UNKNOWN TIME-VARYING PARAMETERS

The initial values of thrust, mass and centre of gravity are used to design the missile's controller, as these will be known at missile launch, however after launch these parameters will vary. The time-varying parameters are used to model the missile dynamics that the controller does not know whilst still being able to control the system. The significant changes in the parameters shown lead to large variations in the dynamics of the missile system, as each impacts a different part of Eq. (11) and Eq. (12). These changes are what the control system must overcome without knowing what they are.

### 4.1 THRUST

The missile is considered during the initial boost and then the cruise phase of flight. Missiles typically use a boost-sustain thrust profile [19]. Cruise thrust is estimated at 5kN, with the boost profile from Won [7]. This produces the profile shown in Figure 5.

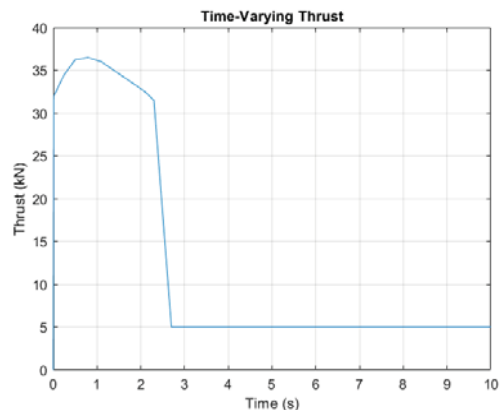
**Figure 5: Missile thrust profile.**

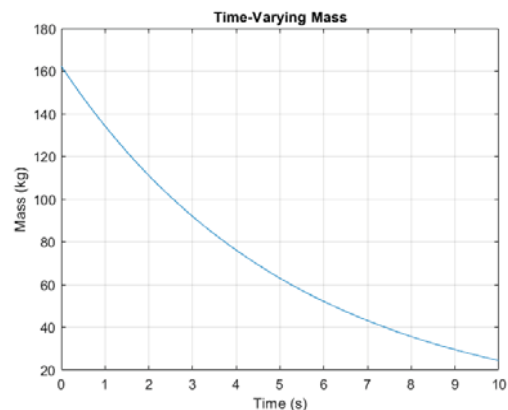
**Figure 6: Missile mass over time.**

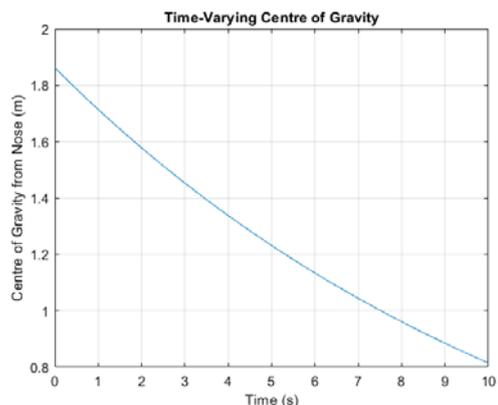
**Figure 7: Centre of gravity distance from missile nose over time.**

## 4.2 MASS

The mass profile is obtained assuming that it loses 40% of its total weight within 3 seconds of launch [8]. The profile used is from Siouris [19]:

$$m(t) = m_0 \left[\frac{m_f}{m_0}\right]^{\frac{(t-t_I)}{t_b}} \quad (13)$$

Where $m_f$ is the final missile mass, $t_I$ is the ignition time, and $t_b$ is the burn time. This produces the profile shown in Figure 6.

## 4.3 CHANGE OF CENTRE OF GRAVITY

As the mass of the missile changes with time, so must the centre of gravity, as the fuel is not uniformly distributed along the missile body and is not burned such that the centre of gravity stays the same. To model the change in centre of gravity we need to find out what the initial ($x_{CG_0}$) and final ($x_{CG_f}$) centres of gravity are.

The initial centre of gravity is taken as being at half the missile's length [20]. The percentage change in the centre of gravity is approximated from work by Won as 20% of the initial centre of gravity [8].

The centre of gravity profile can be based upon the mass profile as:

$$x_{CG}(t) = x_{CG_0} \left[\frac{x_{CG_0}}{x_{CG_f}}\right]^{\frac{(t-t_I)}{t_b}} \quad (14)$$

This produces the profile shown in Figure 7.

## 4.4 MEASUREMENT NOISE EXISTENCE

A missile uses an inertial measurement unit (IMU) to measure accelerations and angular rates and integrate angular rates to obtain positions. A typical IMU that may be used in a missile is the LITIS® offered by Collins Aerospace [21]. This contains their CRS39-03 gyroscopes which have an angle random walk (ARW) of $ARW = 0.015°/\sqrt{hr}$. This number describes the average error that will occur when the signal is integrated. The noise is implemented in the simulation as band-limited white noise with the noise power (NP), Eq. (15), in $rad^2/s$ to match the missile dynamics output. The sample time is set to 200 Hz.

$$NP = \left(ARW \cdot 60 \cdot \frac{\pi}{180}\right)^2 \quad (15)$$

A second-order filter was designed to reduce the noise going to the integrator. This was chosen over a first-order filter as it is better at maintaining the dynamic performance of the system and does not distort the input signal. Optimal performance was achieved at a cut-off frequency of 25 Hz and a damping ratio ($\zeta$) of 0.7.

$$\omega_n = 2\pi \cdot f_{cut-off} \quad (16)$$

$$\frac{\dot{\theta}(s)_{noisy}}{\dot{\theta}(s)_{filtered}} = \frac{\omega_n^2}{s^2 + 2\zeta\omega_n + \omega_n^2} \quad (17)$$

This gives the performance shown in Figure 8. For future improvements in noise reduction an extended Kalman filter may be used, due to its ability to manage the estimation of non-linear system states. The second order filter was chosen instead for this paper as noise filtering was not a main objective but should be considered in the future to improve system performance.

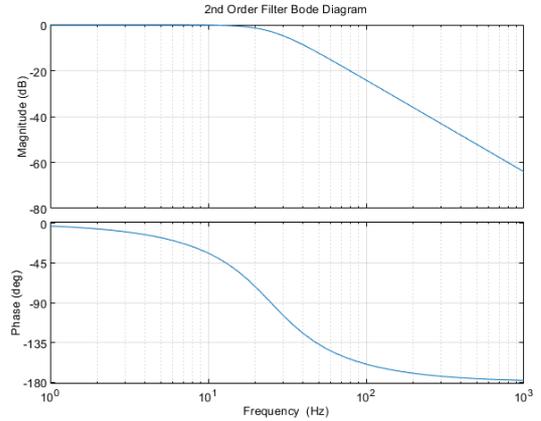

**Figure 8: Bode diagram of the second-order noise filter.**

## 5. MAIN RESULTS

### 5.1 MISSILE CONTROLLER DESIGN WITH CHATTERING REDUCTION

#### 5.1.1 MISSILE CONTROLLER EQUATIONS

The sliding mode controller was designed based upon the pitch dynamics in Eq. (12). Using the definition from Eq. (10), the pitch dynamics become:

$$\dot{q} = f + gu + d \quad (18)$$

Where:

$$f = \frac{QS_{ref}dC_{m\alpha}}{I_{yy}}\alpha \quad (19)$$

$$g = \frac{1}{I_{yy}}\begin{bmatrix} T_0(l - x_{CG}) \\ QS_{ref}dC_{m\delta_T} \end{bmatrix} \quad (20)$$

$$d = \frac{QS_{ref}d\Delta C_{m\alpha}}{I_{yy}}\alpha - \frac{(T_0 + \Delta T_0)\Delta x_{CG}}{I_{yy}}\delta_N$$
$$+ \frac{(T_0 + \Delta T_0)}{I_{yy}}(l\Delta\delta_N - x_{CG}\Delta\delta_N - \Delta x_{CG}\Delta\delta_N) \quad (21)$$
$$+ f(\Delta m_0)$$

Sliding mode controllers, in general, have a problem with a phenomenon known as chattering. This behaviour occurs when the switching function of the controller rapidly changes between one state and the other. Chattering is harmful to actuators as it limits their responsiveness and causes mechanical wear, leading to failures [14]. The conventional switching function is the sign function (Figure 9a), which has a discontinuity at $s = 0$ leading to chattering. To counteract this, the switching function can be changed. In this case the saturation function (Figure 9b) has been selected for its portion of linear control regulated by the boundary layer selection parameter, $\Delta$.

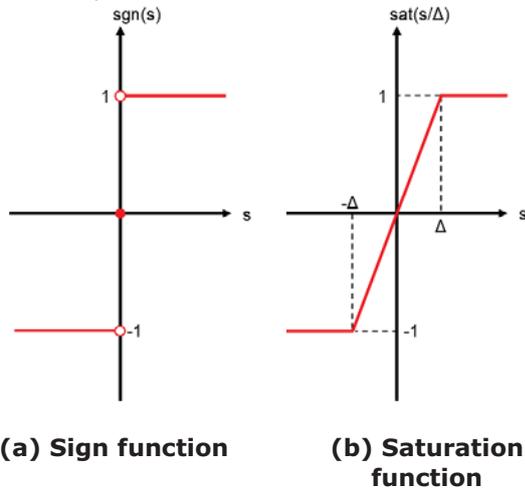

(a) Sign function    (b) Saturation function

**Figure 9: SMC Switching Functions.**

The final controller is then designed as per Eq. (6):

$$u = \frac{1}{g}\left[-f + \ddot{\theta}_d + c\dot{e} + \eta\,\mathrm{sat}(s)\right] \quad (22)$$

To make the system tend to stability, $\eta$ is selected as:

$$\eta = ||d|| \quad (23)$$

### 5.1.2 MISSILE CONTROLLER GAIN TUNING

To achieve the control goals of a time constant of less than 0.35s and a steady-state error of less than 5% the gains of both the elevator controller and thrust vector controller (TVC) needed to be adjusted (Figure 4). The missile system with controllers, actuators, and signal noise was modelled in MATLAB Simulink. Each gain was varied from 0 to 100% effectiveness with steps of 2%. For some combinations of controller gains the simulations did not run successfully due to instabilities in the actuator modelling, hence there are some gaps in the resulting graphs. The step-input for the optimisation was 35°. For a summary of system tuning see Table 3.

**Table 3: Summary of tuned system properties.**

| System Property | Value |
|---|---|
| TVC Gain | 0.32 |
| Elevator Gain | 0.66 |
| Time Constant (s) | 0.16 |
| Settling Time (s) | 0.58 |
| Maximum Overshoot (°) | 1.67 |

Figure 10 highlights the viable selection region for the controller gains. Each system simulation shown in green has a time constant of less than 0.35s which meets the goal of the controller performance from Reichert [1].

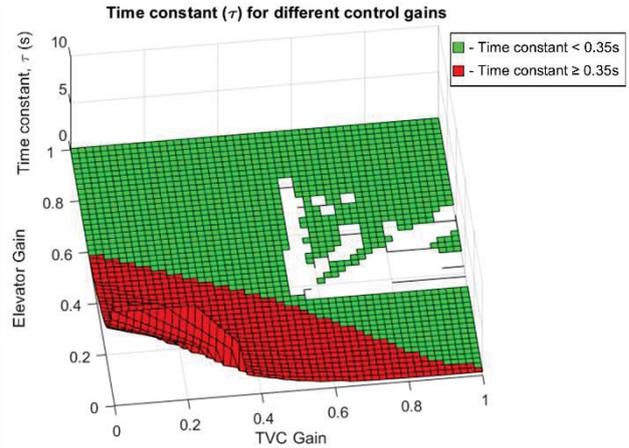

**Figure 10: Time constant variation for varying control gains.**

The next goal to meet is to have a following error of less than 5%. For each of the viable control gains, the time taken for the resulting system to stabilise within 5% of the desired step input was extracted from the simulation and plotted, with the optimum control gains being 0.32 and 0.66 for the TVC and elevator controller respectively, resulting in a settling time of 0.58s (Figure 11). This ensures that the system responds as quickly as possible within the allowable error.

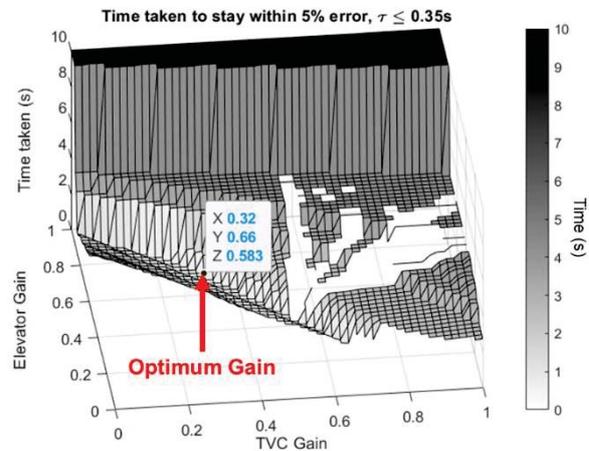

**Figure 11: Time to settle to ±5% of the desired value for varying control gains.**

The resulting surface of maximum overshoot (Figure 12) was examined to ensure that the missile did not go past the required step-input by a large margin and rapidly change direction, as this would impose significant structural stresses on the missile body during flight. The results show only a 1.6° (4.9%) overshoot compared to the 48.4% seen in the robust H∞ controller designed by Mahmood [5].

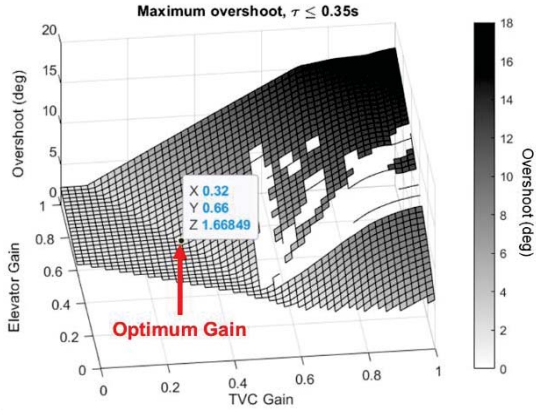

**Figure 12: Maximum overshoot for varying control gains.**

### 5.2 MISSILE ACTUATOR DESIGN WITH DELAY CORRECTION

The current SMC control methods [3, 4] have limited or no models for the actuator dynamics. Actuators need to be considered as they place a significant limit on the controller performance, since it takes time to deflect a control surface to a certain angle and change its direction, i.e., they have bandwidth. This leads to a difference between the desired output and the actual output and hence leads to an increase in system settling time which negatively impacts the desired performance.

The actuators modelled in this paper use values obtained from literature by Wassom [6] for each actuator and are presented in Table 4. The natural frequency is obtained so that the actuators can be modelled in Simulink.

$$\omega_N = \frac{2\pi \cdot f_{BW}}{\sqrt{1 - 2\zeta^2 + \sqrt{2 - 4\zeta^2 + 4\zeta^4}}} \quad (24)$$

Where $\omega_N$ is the natural frequency, $f_{BW}$ is the bandwidth, and $\zeta$ is the damping ratio.

**Table 4: Actuator performance limits.**

| Requirement | Nozzle actuator | Elevator actuator |
|---|---|---|
| Maximum deflection (°) | 15 | 30 |
| Bandwidth (Hz) | 40 | 45 |
| Slew rate (°/s) | 400 | 600 |
| Damping ratio | 0.3 | 0.3 |
| Natural frequency (rad/s) | 173 | 195 |

The actuator bandwidth limitation leads to a response lag in the system, therefore a phase lead compensator can be included to mitigate these effects [10] and increase actuator responsiveness. These are based upon the time-constant (T) of the respective actuator, given by:

$$T = \frac{1}{\zeta \cdot \omega_N} \quad (25)$$

The compensator can then be designed in the frequency domain as:

$$\frac{\delta_{compensated}(s)}{\delta_{reqired}(s)} = \frac{Ts + 1}{nTs + 1} \quad (26)$$

Where $n$ is the compensator parameter to be selected. The effect of not including and including the compensator can be seen in Figure 13 and Figure 14 respectively. Without the compensator the system can better reject system noise, as seen from the smoother curve in Figure 13, however eventually the actuator eventually cannot keep up with the bandwidth requirement of the controller, leading to large instabilities at 7 seconds.

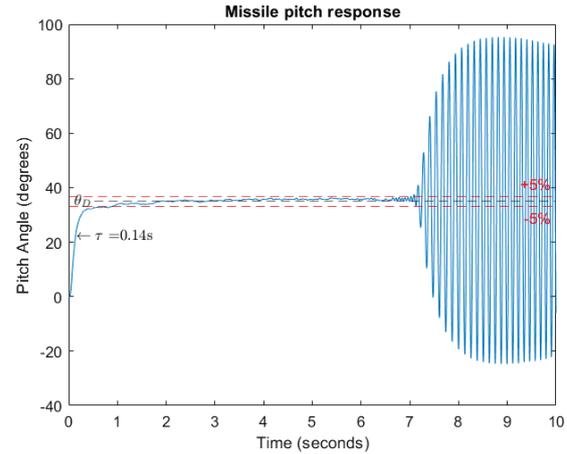

**Figure 13: Missile pitch response with gyroscope noise, no phase compensator.**

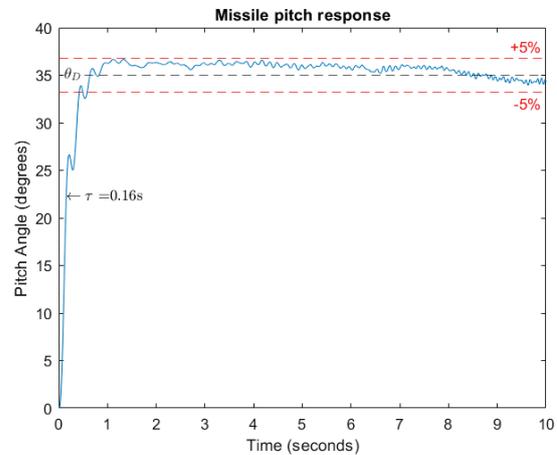

**Figure 14: Missile pitch response with gyroscope noise and a phase compensator.**

With the addition of the phase compensator some additional noise is introduced from the control system switching, however it is limited in what bandwidth it can pass to the actuators, and they are able to keep up with their requirements.

The detailed actuator performance that has been calculated improves upon the basic methods used in previous work [3, 4] to represent the controller performance more realistically in the full missile control loop.

### 5.3 SIMULATION RESULTS AND ANALYSIS

Using the tuned SMC system from Section 5.1 the controller was verified to successfully reject the disturbances outlined in Section 4, working across a range of desired pitch angles, 0° to 90° (Figure 15a). Similar behaviour is seen for all the required step-input angles.

#### 5.3.1 ANALYSIS ON RISE TIME

For all required input angles, the rise time remains under the required 0.35s (Figure 15b). The non-linear adaptive SMC designed by Moon and Lee manages to achieve rise times of less than 0.2s without excessive overshoot [4], whereas for the proposed SMC only angles under 60° achieve rise times of under 0.2s, with the rise time increasing generally with desired pitch angle. For future SMC controllers it therefore may be desirable to consider using the adaptive sliding gain proposed by Moon and Lee [4].

#### 5.3.2 ANALYSIS ON SETTLING TIME

The settling time stays around same value for desired angles over 40° but rapidly increases below that value (Figure 15c). This is due to the 5% error margin becoming smaller as the target value becomes smaller, making it more difficult for the controller to achieve the desired value in a short amount of time. To counter this there may need to be an optimisation of control gains for a variety of desired pitch angles, however this strays towards the gain-scheduling that the SMC was aiming to avoid. If larger errors are acceptable for smaller angles, then this may not be a problem as presently, for example, for 10° the allowable error is only ±0.5°.

#### 5.3.3 SYSTEM IMPROVEMENTS

The system performance is severely limited by the presence of gyroscope noise. Figure 16 highlights the difference between what the gyroscope measures as the system error, including noise filtering, and the true value for the system error. This difference shows that the SMC is successful in controlling

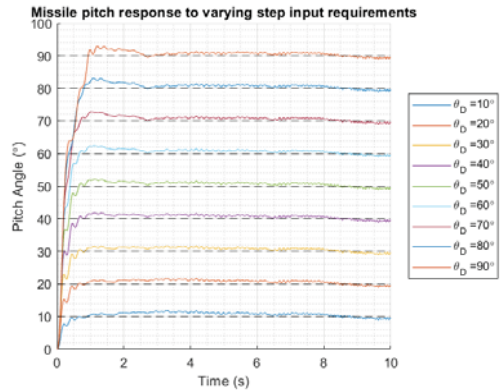

**(15a) Pitch angle response versus time.**

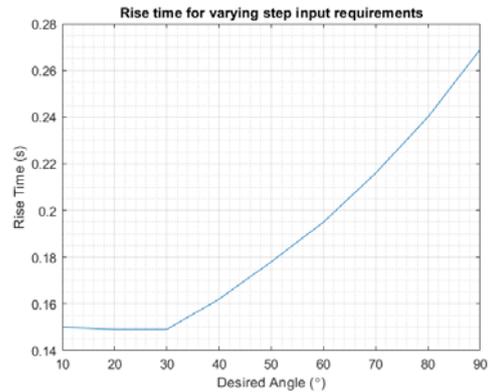

**(15b) Rise time versus required angle.**

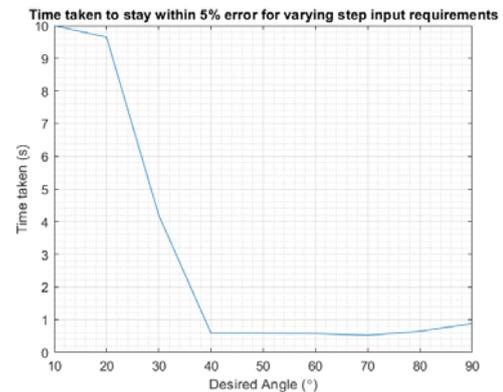

**(15c) Time to settle to ±5% of the desired value versus required angle.**

**Figure 15: System responses to different desired pitch angle requirements.**

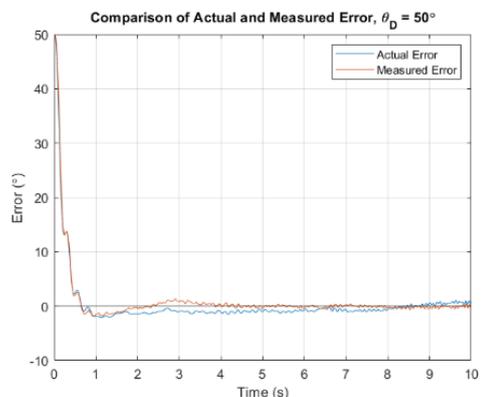

**Figure 16: Comparison of measured and actual error for a desired pitch angle of 50°.**

the measured error to 0°, but the limit imposed by the gyroscopes means that it is not controlling the true error. The difference in the errors accumulates over time due to the integration of gyroscope angular rate to obtain angular position. This could be solved with improved gyroscope noise filtering.

## 6. CONCLUSIONS

The designed robust control system based on the saturated sliding mode controller was successful at achieving the time constant of less than 0.35s and a steady-state error of less than 5% for a range of desired pitch angles, despite experiencing rapid changes in the thrust, centre of gravity, and mass of the missile. Furthermore, the sensor measurement noise and actuator requirements do not limit its effectiveness because of the use of the second-order filter and phase compensator.

The effects of wind disturbances were not simulated, and the missile was only considered in 1-DoF (pitch rotation).

Future work should consider a full atmospheric model, implementation the control loop SMC within a guidance loop context and investigate improved noise filtering techniques to reduce the difference between actual and measured error.